\documentclass[11pt,twoside]{article}
\usepackage{asp2010}

\resetcounters

\begin{document}

\title{The Local Group: Inventory and History}
\author{Eline Tolstoy$^1$
\affil{$^1$Kapteyn Astronomical Institute, University of Groningen, the Netherlands}}

\begin{abstract}
My presentation was an overview of what we know about the Local Group
of galaxies, primarily from optical imaging and spectroscopy.  AGB
stars are on the whole a very sparse and unrepresentative stellar
population in most Local Group galaxies.  However, more detailed
studies of star formation histories and chemical evolution properties
of populations, like Main Sequence dwarf stars and Red Giant Branch
stars, allow a better understanding of the evolutionary context in
which AGB stars can be observed.  There are a variety of galaxy types
in the Local Group which range from predominantly metal poor (e.g.,
Leo~A) to metal-rich (e.g., M~32).  Dwarf galaxies are the most
numerous type of galaxy in the Local Group, and provide the
opportunity to study a relatively simple, typically metal-poor,
environment that is likely similar to the conditions in the early
history of all galaxies.  Hopefully the range of star formation
histories, peak star formation rates and metallicities will provide
enough information to properly calibrate observations of AGB stars in
more distant systems, and indeed in integrated spectra.  Here I will
summarise what we know about the star formation histories of nearby
galaxies and their chemical evolution histories and then attempt to
make a connection to their AGB star properties.
\end{abstract}

\section{Introduction}

Within the Local Universe galaxies can be studied in great detail star
by star. The Colour-Magnitude Diagram synthesis analysis method is
well established, at optical wavelengths, as the most accurate way to
determine the detailed star formation history of galaxies going back
to the earliest times \citep[e.g.,][]{Tolstoy09}.  This approach has
benefited enormously from the exceptional data sets that wide field
CCD imagers on the ground and the Hubble Space Telescope can
provide. Spectroscopic studies using large ground-based telescopes
have allowed the determination of abundances and kinematics for
significant samples of red giant branch (RGB) stars and also more
massive O, B and A stars in several nearby galaxies \citep[e.g.,][and
references therein]{Tolstoy09}. These studies have shown directly how
properties can vary spatially and temporally, and how this information
can give important constraints to theories of galaxy formation and
evolution.

\begin{figure}
\hskip 1.5cm
\resizebox{0.85\columnwidth}{!}{\includegraphics{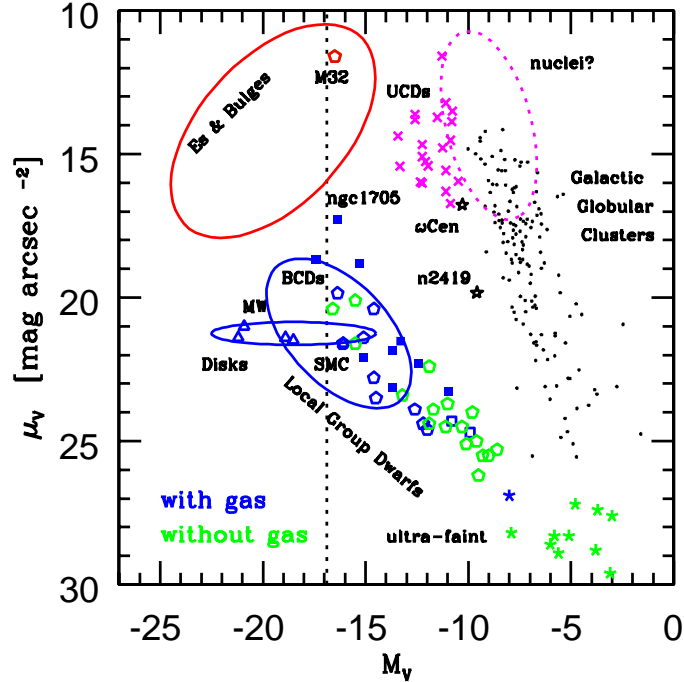}}
\vskip -0.7cm
\caption{
The relationship between the structural properties (absolute
magnitude, M$_V$ and central surface brightness $\mu_V$) for a range
of different galaxy types. The dotted line is the classical maximum
luminosity of the dwarf galaxy class, from Tammann (1994). Local Group
galaxies are plotted as open pentagons, with the colour depending upon
their gas content. The Sloan discovered ultra-faint systems as plotted
as star symbols.  Blue Compact Dwarf galaxies are squares,
Ultra-compact systems as crosses and Galactic globular clusters as
dots.  See \citet{Tolstoy09} and \citet{Binggeli94} for more details.
}
\label{bing}       % Give a unique label
\end{figure}

\begin{figure}
\hskip 1.5cm
\resizebox{0.85\columnwidth}{!}{\includegraphics{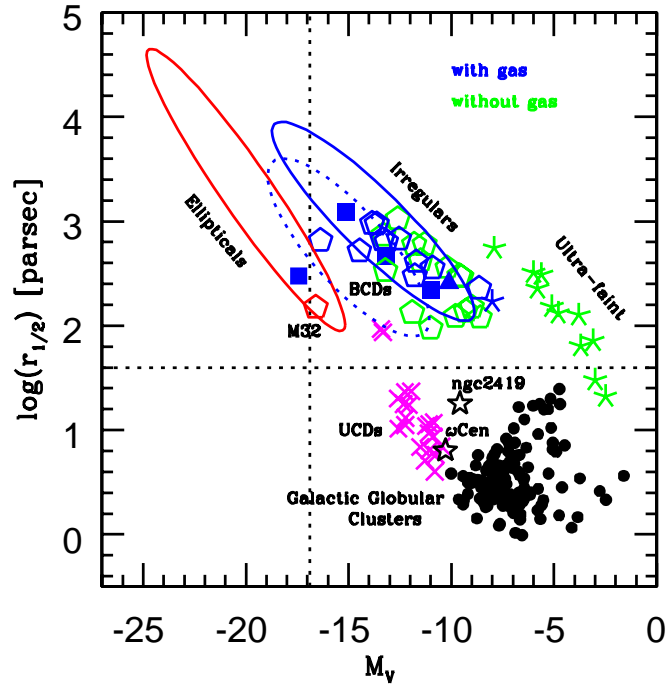}}
\vskip -0.7cm
\caption{
The relationship between the structural properties (absolute
magnitude, M$_V$ and half-light radius $r_h$) for a range
of different galaxy types. The dotted lines are the classical maximum
luminosity of the dwarf galaxy class, and the minimum spatial extent, 
from \citet{Tammann94}. The symbols are the same as in Figure~\ref{bing}.
See \citet{Tolstoy09} and \citet{Belokurov07} for more details.
}
\label{bel}       % Give a unique label
\end{figure}

Dwarf galaxies are commonly used as probes of a simple ``single cell''
star forming environment. They cover a range of mass and metallicity,
and are considered to be representative of how galaxies in the early
universe may have looked.  A working definition of dwarf galaxies
includes all galaxies that are fainter than M$_B \le -16$ (M$_V \le
-17$) and more spatially extended than globular clusters
\citep[e.g.,][]{Tammann94}, see Figures~\ref{bing}~\&~\ref{bel}.
Although these limits were not physically motivated, they are broadly
consistent with the limit of mass and concentration at which gas
outflows are likely to start to significantly effect the baryonic mass
of a galaxy. This includes a number of different types: early-type
dwarf spheroidals (dSphs); late-type star-forming dwarf irregulars
(dIs); the recently discovered very low surface brightness,
ultra-faint, dwarfs (uFd); as well as centrally concentrated actively
star-forming blue compact dwarf galaxies (BCDs). The newly discovered,
even more extreme, so-called ultra-compact dwarfs (UCDs) are
identified as dwarf galaxies from spectra but are of a similar
compactness to globular clusters (see purple crosses in
Figure~\ref{bel}).  The dIs, BCDs, dSphs, late-type and spheroidal
galaxies tend to overlap with each other in global properties in
Figures~\ref{bing}~\&~\ref{bel}. These overlapping properties of early
and late-type dwarfs have long been assumed to be convincing evidence
that early-type dwarfs are late-type systems that have been stripped
of, or otherwise used up, their gas \citep[e.g.,][]{Kormendy85}.

Thus, like larger systems, the global properties of dwarf galaxies
correlate closely with luminosity, half-light radius and surface
brightness, over a large range.  Dwarf galaxies thus allow us to study
specific aspects of galaxy formation and evolution on a small scale.

\section{Optical Imaging: Star Formation Histories}

\begin{figure}
\resizebox{\hsize}{!}{\includegraphics{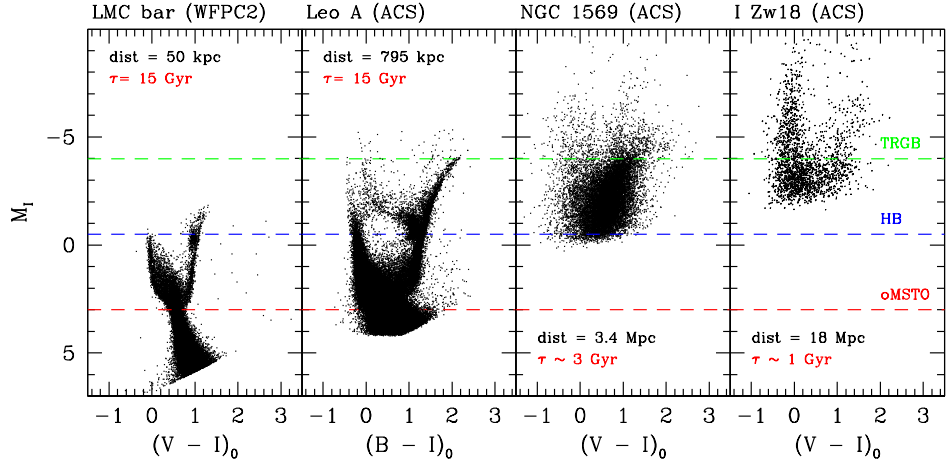}}
\vskip -0.3cm
\caption{ 
The effect of distance on the resolution of individual stars and on
the corresponding look-back time, $\tau$, of the star formation
history.  The CMDs are in absolute magnitude (M$_I$) and colour of
systems all observed for long exposure times with the HST and
photometered with the same techniques, but at different distances. The
LMC bar (50~kpc) \citet{Smecker02}; Leo~A (795~kpc) \citet{Cole07};
NGC~1569 (3.4~Mpc) \citet{Grocholski08, McQuinn10}; I~Zw18 (18Mpc)
\citet{Aloisi07}.
}
\label{cig}       % Give a unique label
\end{figure}

There are increasingly significant difficulties in obtaining and
accurately interpreting the CMDs of galaxies at distances beyond the
Local Group, see Fig.~\ref{cig}.  It is only possible to observe
galaxies star by star in the very nearby Universe (predominantly
within the Local Group), meaning that there are selection effects that
will almost certainly bias our conclusions from these types of
studies.  The main uncertainty is due to the fact that we can only
study the star formation history (SFH) back to the earliest times
within the halo of the Milky Way and in very nearby galaxies, see
Figure~\ref{cig} and also \citet{Cignoni10}.  These galaxies have most
likely suffered significant evolutionary effects, as suggested by the
morphology-density relation \citep[e.g.,][]{Mateo08}. It will be hard
to remove this bias in our observations until a significant leap in
sensitivity and resolution can be made to allow us to look to greater
distances with comparable accuracy (e.g., a large space telescope or
an extremely large ground based telescope working near to its
diffraction limit).

Main Sequence star luminosities have a clear age dependence, and are
thus by far the most accurate age indicators of a resolved stellar
populations as part of the full fitting of the colour-magnitude
diagram \citep[e.g.,][]{Aparicio04}.  The fact that a stellar
population is resolved down to the oldest main sequence turnoffs (M$_I
\sim 3$) means that the luminosity bias that is so apparent in
integrated light studies can be largely removed.  A significant amount
of effort has gone into this kind of work from both large format CCD
observations of very nearby galaxies \citep[e.g.,][de Boer et al., in
prep.]{Hurley98, Harris09}, which are large on the sky and also from deep HST
observations for more distant systems \citep[e.g.,][]{Skillman03,
Cole07, Monelli10} .

Because the number and range of galaxy types in the Local Group is
strongly biased to dwarf galaxies, this is the main type of galaxy
studied with this detail.  Dwarf galaxies are also more straight
forward to observe a large fraction of the system in ``one shot'' even
with HST. There have been numerous detailed studies of individual
dwarf galaxies \citep[e.g.,][and references there in]{Tolstoy09}.
There has also been a project to treat uniformly a large set of
archival HST WFPC2 observations of Local Group galaxies, and create
accurate star formation histories in a consistent manner
\citep{Dolphin05, Holtzman06}.  There have also been challenging
studies of compact systems with extreme crowding, like M~32
\citep{Monachesi10}, backed up by RR Lyr studies
\citep{Fiorentino10}. There have also been deep observations of small
HST fields in the M~31 halo \citep[e.g.,][]{Brown03} and LMC
\citep[e.g.,][]{Holtzman99, Smecker02}

To look at currently more actively star forming systems, for example
Blue Compact Dwarfs we need to look beyond the Local Group
\citep[e.g., NGC 1569 at 3.4~Mpc, see][]{Grocholski08, McQuinn10}, see
Fig.~\ref{cig}. As we get more distant, it becomes harder to detect
anything other than bright stars, and the photometric errors tend to
smear out the features of the CMD. Going from left to right in
Figure~\ref{cig} it can be seen that the features in the CMDs become
less and less well defined. This is mostly due to photometric errors
due to the increasing faintness of the stars, but the related effect
of increasing crowding, that makes it difficult to accurately
disentangle the measurements of (faint) individual stars from their
neighbours.  Often there are clearly a large number of stars present
above the tip of the RGB in BCD galaxies 
(e.g., NGC~1569). These may be either the
effects of crowding, or they may indicate the presence of AGB stars and
that a significant amount of star formation has occurred a few Gyr
ago.

\begin{figure}
%\hskip 1cm
\resizebox{\hsize}{!}{\includegraphics{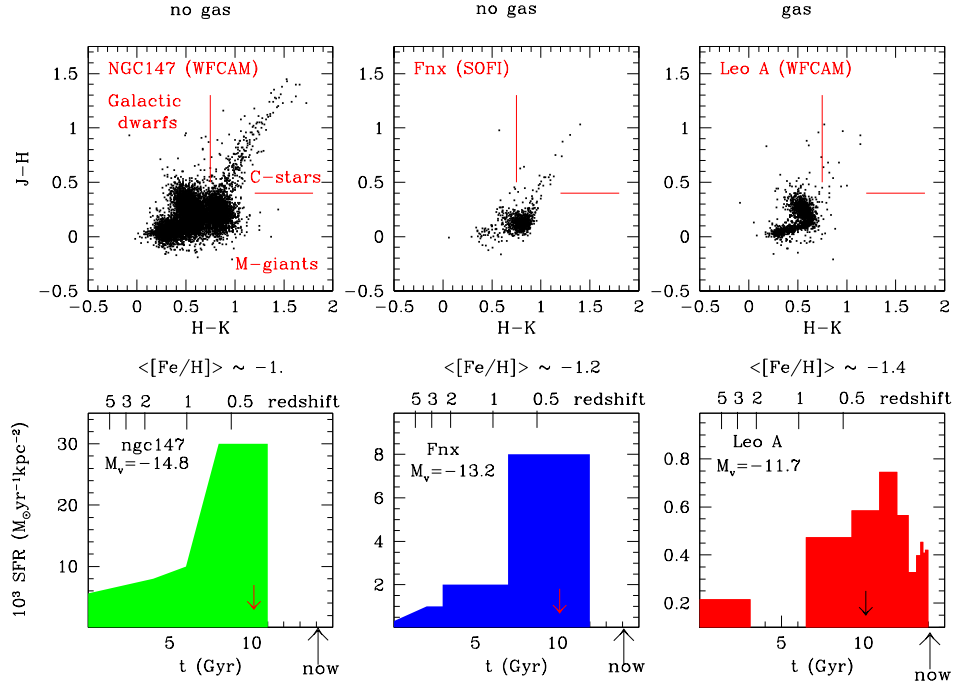}}
\vskip -0.3cm
\caption{ 
In the top panels are the Infra-Red colour-colour diagrams for three
dwarf galaxies: NGC147 (WFCAM data, over 0.8 sq deg, Irwin et al. in
prep), Fornax dSph \citep[SOFI data, over 0.1 sq deg,][]{Gulli07} \& LeoA (WFCAM data, over 0.8 sq deg, Irwin et al. in
prep). The different stellar evolution phases that are delineated in
an accurate colour-colour diagram sequences are labeled in the left
most diagram. The C-stars (or AGB stars) and the M-giants are in the
galaxy itself. The rest of the stars are predominantly Galactic dwarf
stars.  In the lower panels are the corresponding star formation
histories for the same three galaxies, from \citet{Dolphin05}
(NGC~147); \citet{Tolstoy01} (Fnx) and \citet{Cole07} (Leo A).
}
\label{sfh}       % Give a unique label
\end{figure}

Of course this difficulty in detecting faint (blue) main sequence
turnoff stars may have an obvious alternative in the presence of
bright red AGB stars in imaging of galaxies extending to distances
well beyond the Local Group \citep[e.g.,][]{Girardi10}.  However,
without a better calibration of the effects of age and metallicity on
the AGB population it is hard to quantify their presence in terms of
an accurate star formation rate at a given time \citep[e.g., see also
VII~Zw403,][]{Lynds98}.  The CMDs in Figure~\ref{cig} do not always
give a good overview of the very red stars, such as AGB stars in these
galaxies. This is because they do not always stand out very clearly in
optical CMDs. What is really needed are infra-red observations of this
population, and colour-colour diagrams can be especially useful
\citep[e.g.,][]{Cioni03, Gulli07}, see Figure~\ref{sfh}. These populations
can then be calibrated in terms of ages and metallicities coming from
optical imaging and spectroscopy.

In Fig.~\ref{sfh} we look at the star formation histories and also the
number of AGB stars (as seen in IR colour-colour diagrams) 
in three nearby galaxies with a range of
luminosity (from M$_v = -14.8 \rightarrow -11.7$) and also a range of mean
metallicity ([Fe/H] $= -1 \rightarrow -1.4$) at the time the AGB stars
were born. These three galaxies (NGC~147, Fornax dSph \& Leo~A) were chosen
because they have very similar star formation histories, with a peak 
around 3$-$5~Gyrs. In each case the absolute rate at the peak is 
quite different. The more luminous the
galaxy, the higher the peak star formation rate.  But they all had
their peak activities at a similar period in the past.  
The number of C-stars (AGB stars) that
can be seen in the colour-colour diagrams varies by a larger amount
than the SFR differences might imply, especially in the case of
NGC~147.  This might suggest that an important factor is also the
metallicity at which the stars were forming 3$-$5~Gyr ago (these are
also labelled in Fig.~\ref{sfh}). The C-stars in Leo~A still need to
be carefully studied. These would likely be the most metal poor
C-stars in the Local Group, if confirmed, given that the present day
HII region abundance is a mere 3\% of solar \citep{Vanzee06}. The
stars in the C-star region of the colour-colour diagram for Leo~A look
more untidy than the usual AGB sequence, and my well be the result of
confusion or young (massive) stars in HII region.

When you look at the SFHs of dwarf galaxies as a group there is no
discernible trend in either duration or average age of stellar
population with either mass, luminosity or rotation, they seem to
reach a similar luminosity by distinct routes
\citep[e.g.,][]{Skillman07}.  The only effect seems to be that when a
galaxy forms stars, everything else being equal, the maximum rate
seems proportional to the mass of the galaxy, that is to the total
number of stars formed, but not when they formed.  How the number of
evolved stars (e.g., carbon stars, or E-AGB stars) fits into SFH has
not yet been clearly quantified.  The number of AGB stars should be
studied for a range of galaxies using the accurate SFHs from deep
optical data where available to better understand if it is possible to
disentangle the effects of age, metallicity and small number
statistics in the interpretation of their properties.

\section{Optical Spectroscopy: Abundance Properties}

\begin{figure}
\resizebox{\hsize}{!}{\includegraphics{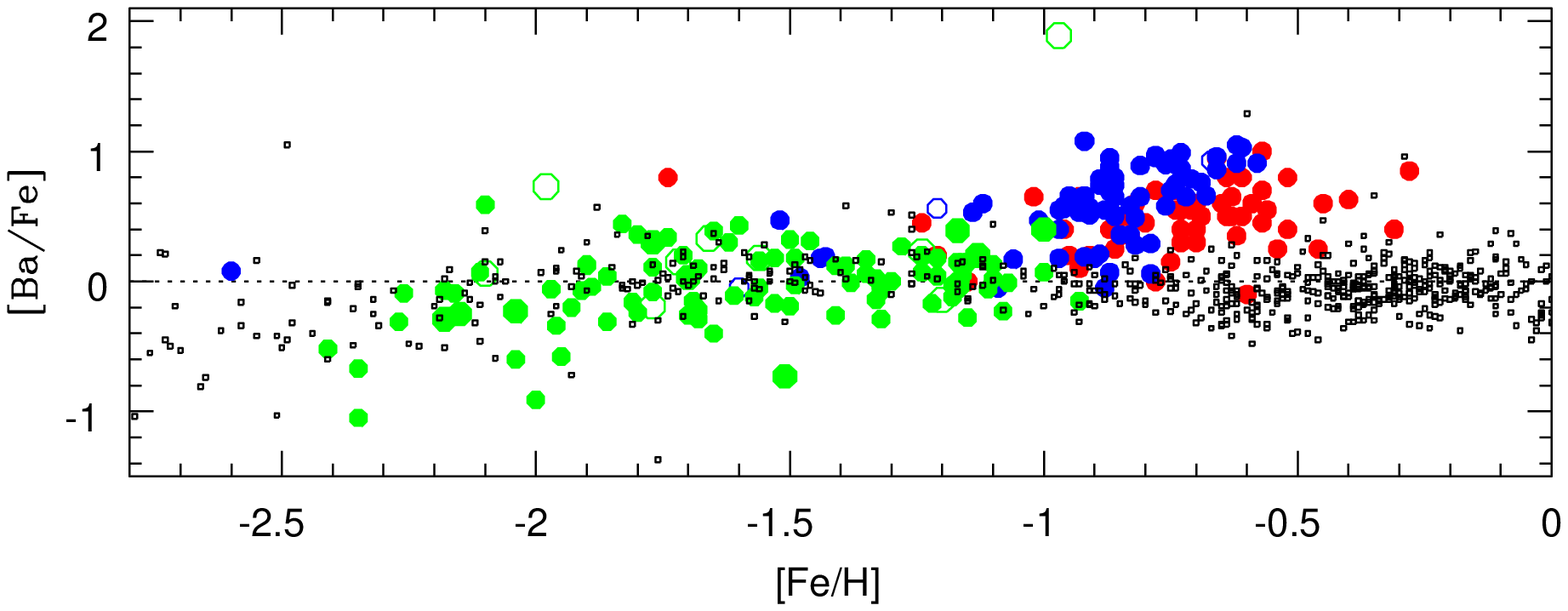}}
\vskip -0.3cm
\caption{ 
Here are the high resolution abundances of individual red giant branch
stars in Sculptor dSph (green solid circles, Hill et al. (2011) in prep, FLAMES
high resolution; open circles, \citet{Shetrone03}, UVES); Fornax dSph
(blue solid circles, \citet{Letarte10}, FLAMES high resolution; open
circles, \citet{Shetrone03}, UVES); Large Magellanic Cloud (red
circles, \citet{Pompeia08}, FLAMES high resolution).  The small
black squares are Galactic observations \citep[from compilation,][]{Venn04}.
}
\label{bar}       % Give a unique label
\end{figure}

For the most galaxies in the Local Group it is possible to take
spectra of large samples of individual RGB stars at intermediate
resolution. This allows the observation of well calibrated, simple to
use metallicity indicators, such as the Ca~II triplet
\citep[e.g.,][]{Starkenburg10, Battaglia08cat}.  These measurements
allow a detailed measurement of the metallicity distribution function
from many hundreds and sometimes even thousands of individual stars.
The kinematic properties of galaxies, can also be disentangled with
these spectra \citep[e.g.,][]{Battaglia08mass}, as well as any
connection between distinct kinematic components and metallicity. This
leads to accurate mass modelling of individual galaxies and also to
the discovery of distinct kinematic components, even in small dwarf
galaxies, and sometimes also rotation \citep[e.g.,][]{Lewis07,
Fraternali09}.

In the most nearby systems (ie., mostly dwarf galaxies, but also the
Magellanic Clouds) it is possible to take high resolution spectra of
individual RGB stars.  This allows us to measure detailed abundances
of numerous chemical elements. The most commonly observed are alpha
elements (e.g., O, Ca, Mg, Ti), but also heavy elements, such as
r-process elements (e.g., Eu), Iron-peak elements (e.g., Mn, Cr, Fe,
Ni) and also s-process elements (e.g., Ba). The abundances of these
elements in RGB stars allow us to probe their levels over the entire
star formation history that occurred $>$~1~Gyr ago.  This allows us to
follow which enrichment processes dominate at different epochs in the
galaxy, and thus their time scale, and how they effect and are
effected by the presence or absence of other elements.  

The most important elements for tracing the effect of AGB stars and
their pollution of the ISM out of which subsequent generations of
stars are made are s-process elements.  Fig.~\ref{bar} shows the
detailed abundances of Barium compared to Iron, [Ba/Fe], based on high
resolution spectroscopic observations of individual RGB stars in the
Sculptor dSph, the Fornax dSph and the Large Magellanic Cloud,
compared to RGB stars in the Galactic disk and halo.  Barium is of
particular interest because at these [Fe/H] values it is produced
almost entirely by the s-process. This also makes it a good indicator
of how many potential s-process sources there have been and when they
were most productive. Fig.~\ref{bar} shows that both the LMC and
Fornax have significantly enhanced [Ba/Fe] compared the Galaxy at
[Fe/H]$> -1$. It seems that this enhancement only starts at
[Fe/H]$\sim -1$. Sculptor thus does not show the same effect,
presumably because it never reached a high enough metallicity before
all star formation stopped. It might also be because Sculptor stopped
forming stars before the feedback of s-process elements from AGB stars
became important to the chemical enrichment.

\begin{figure}
\resizebox{\hsize}{!}{\includegraphics{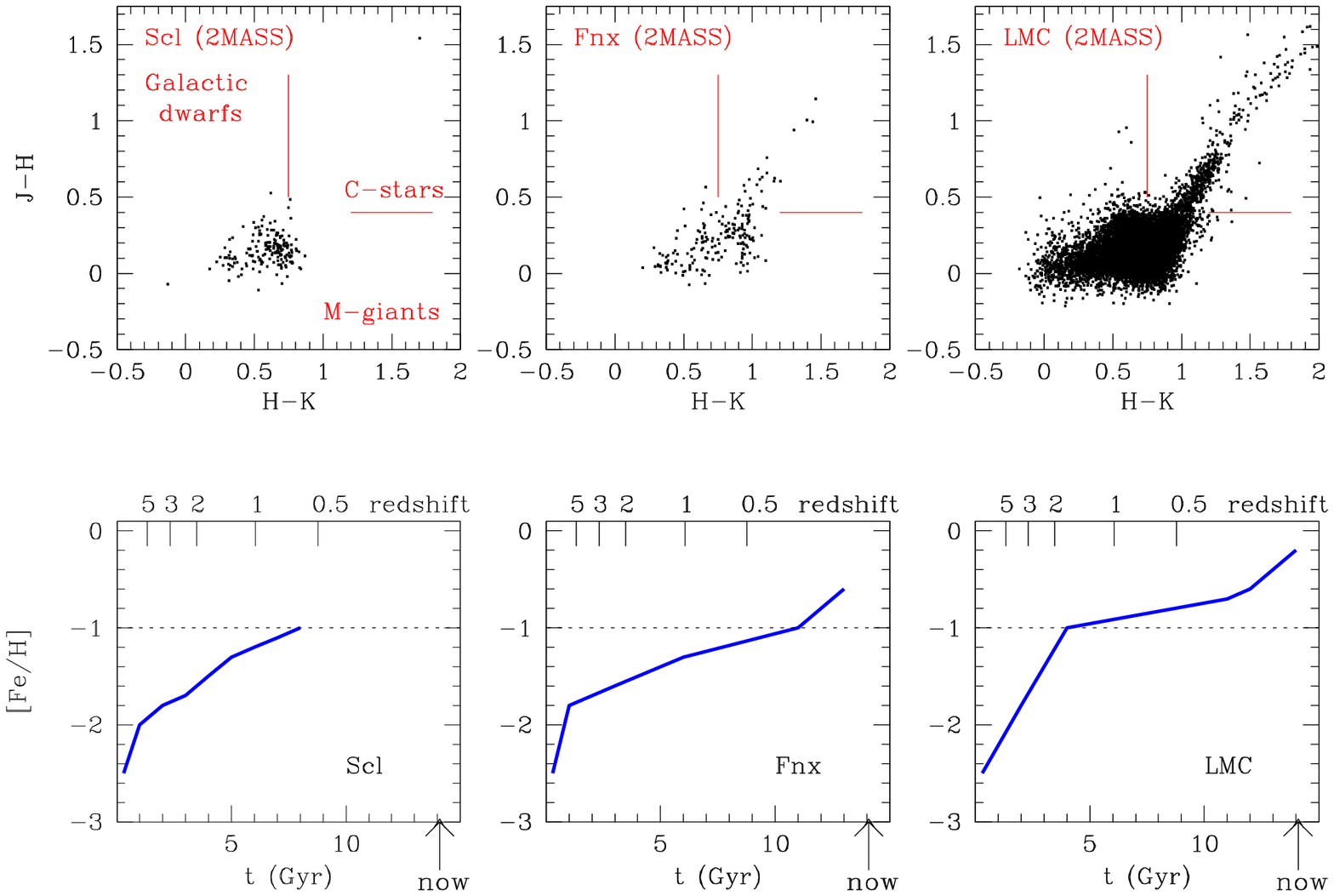}}
\vskip -0.3cm
\caption{ 
For the same galaxies shown in Fig.~\ref{bar} the metallicity-age
relations are shown, as are the colour-colour diagrams which clearly
show the numbers of AGB (C-stars) present. The age-metallicity
relations come from de Boer et al., in prep for Sculptor; \citet{Battaglia06} 
for Fornax and \citet{Pagel98, Hill00} for LMC. The Infra-red
data are all selected from 2MASS (only those stars with AAA quality
flags), in a region that corresponds to the tidal radius of Scl and
Fnx, and within the central 1 degree of the LMC. 
}
\label{zfh}       % Give a unique label
\end{figure}

In Fig.~\ref{zfh} we consider the evolution of [Fe/H] in the same
galaxies shown in Figure~\ref{sfh}, i.e., Sculptor dSph, Fornax dSph
and the Large Magellanic Cloud.  In Fig.~\ref{sfh} we show
colour-colour diagrams coming from 2MASS data for each of the
galaxies. The physical region sampled is the same for Sculptor \&
Fornax (1~deg; which is about the distance to the tidal radius). This
region is a smaller fraction of the whole galaxy for the LMC. Clearly
the LMC is a larger, more luminous (with a higher peak star formation
rate) galaxy than the other two, and the LMC also contains many more
AGB stars.

The variation in the number of AGB stars seen in these nearby galaxies may
be due to the different masses, sizes and/or luminosities of the systems,
but there is also likely to be a significant effect due to
metallicity. It can be seen that the galaxy that never forms stars
with [Fe/H]~$>~-1$ (e.g., Sculptor, see Fig.~\ref{zfh}) also appears
to contain no AGB (C-stars) and no sign of enrichment by these stars
during its star formation history (e.g., Fig.~\ref{bar}). 
Of course Sculptor also stopped
forming stars around 6~Gyrs ago, and for several Gyr before this it
formed stars at a very low rate (see de Boer et al., in prep), thus it
might be a case of low number statistics. But Leo~A is a galaxy with a
similar luminosity to Sculptor, and a current metallicity (from H~II
region spectroscopy) which is similar to the average metallicity found
in Sculptor. Leo~A also formed most of its stars over the last 5~Gyrs and
yet there are very few, if any, AGB in Leo~A (see Fig.~\ref{sfh}).

\section{Conclusions}

It is clear the that AGB stars can play a very significant role in the
chemical evolution of a galaxy, especially a dwarf galaxy.  A dwarf
galaxy with an extended star formation history will likely be highly
sensitive to the chemical enrichment created by the relatively slow
and steady stellar winds from AGB stars. In small galaxies Supernovae
may drive mass and metals entirely out of the galaxy, but stellar
winds from AGB stars probably will not.  The effect of AGB stars is
likely to be dependent upon the time scale over which star formation
occurred. The products of these stellar winds must be returned to the
ISM on a time frame consistent with the subsequent star formation
episodes in a galaxy to have an impact on the chemical evolution.
From the lack of AGB stars in very metal poor systems it also seems
likely that the [Fe/H] plays a role in the evolution of AGB star
populations. It seems to be more difficult to produce metal poor AGB
stars, and also to measure any effect in the abundance ratios that may
come from them. 

In this review I have just touched upon the connections that can be
made between the AGB star properties of nearby galaxies and their star
formation histories and metallicities. These results are likely to be
placed on much more quantitative basis in the coming years as more and
more wide-field near-IR and optical imaging and spectroscopic surveys
are carried out for both nearby and more distant galaxies.  It is
clear that to sort out the complex and intertwined effects of star
formation, stellar winds, supernovae explosions and their effect on
the ISM we need to use information from a variety of sources that are
sensitive to different time scales, and different physical processes.
This means that we need to combine information from optical imaging
(SFHs) and spectroscopy (abundances) with IR imaging and spectroscopy
to get the full story.

\acknowledgements I would like to thank Mike Irwin for useful
conversations and letting me use his unpublished IR data. I would also
like to thank the organisers for inviting me to this most interesting
meeting, and NWO for funding my trip, through a VICI grant.

\bibliographystyle{asp2010}
\bibliography{et}

\end{document}